\documentclass{article}
\usepackage{graphicx}
\usepackage[english]{babel}
\usepackage[margin=1in]{geometry}
\usepackage[ruled]{algorithm2e}
\usepackage{multicol}
\usepackage{subcaption}
\usepackage[square,sort,comma,numbers]{natbib}

\author{S. Chatterjee \footnote{\texttt{rs171@iiita.ac.in}} \and B.S. Sanjeev \footnote{\texttt{sanjeev@iiita.ac.in}} }
\title{
{\bfseries\Large Network-based community detection of comorbidities and their association with SARS-CoV-2 virus during COVID-19 pathogenesis\bigskip}
}

\date{May 31, 2022}

\begin{document}
\maketitle

\begin{center}
Department of Applied Sciences\\ Indian Institute of Information Technology\\ Allahabad 211012, India \\
\end{center}


%

\tableofcontents


\begin{abstract}

Recent studies emphasized the necessity to identify key (human) biological processes and pathways targeted by the Coronaviridae family of viruses, especially SARS-CoV-2. COVID-19 caused up to 33-55\% death rates in COVID-19 patients with malignant neoplasms and Alzheimer's disease. Given this scenario, we identified biological processes and pathways which are most likely affected by COVID-19. The associations between various diseases and human genes known to interact with viruses from Coronaviridae family were obtained from the IntAct COVID-19 data set annotated with DisGeNET data. We constructed the disease-gene network to identify genes that are involved in various comorbid diseased states. Communities from the disease-gene network through Louvain method were identified and functional enrichment through over-representation analysis methodology was used to discover significant biological processes and pathways shared between COVID-19 and other diseases. The IntAct COVID-19 data set comprised of 828 human genes and 10,473 diseases that together constituted nodes in the disease-gene network. Each of the 70,210 edges connects a human gene with an associated disease. The top 10 genes linked to most number of diseases were VEGFA, BCL2, CTNNB1, ALB, COX2, AGT, HLA-A, HMOX1, FGT2 and COMT. The most vulnerable group of patients thus discovered had comorbid conditions such as carcinomas, malignant neoplasms and Alzheimer's disease. Finally, we identified 37 potentially useful biological processes and pathways for improved therapies.


\end{abstract}

\section{Introduction}
\label{intro}

Genes and proteins perform a variety of functions within an organism. Interactions between them are indispensable to fundamental cellular processes. Graph-theoretic tools enable processing network topological information to identify and analyze functional dependencies between disease-gene dynamics \citep{li2020network}. Many network-based strategies have been widely adopted to investigate such relationships \citep{wang2011network}. The current pandemic, COVID-19, is caused by SARS-CoV-2 virus. This virus is a member of Coronaviridae family that is also responsible for other life threatening diseases such as SARS-CoV and MERS \citep{abdelrahman2020comparative}. Insights from disease-gene networks have been useful to decipher the underlying molecular interactions with  context-specific studies, including COVID-19 \citep{li2021network}. Evidence from multiple studies has shown that the severity of COVID-19 is enhanced by comorbid and life-threatening diseases, and patients with comorbidities have been affected disproportionately \citep{zong2021intersection, sanyaolu2020comorbidity}. Such patients are more susceptible to organ failure and mortality \citep{sanyaolu2020comorbidity}. Patients with comorbidities such as carcinomas and malignant neoplasms have a higher risk of developing multiple organ failure leading to death \citep{mehta2020covid}. According to a recent study on COVID-19 patients, the mortality rate was as high as 33\% if afflicted with malignant neoplasms \citep{chavez2022evaluation}. At 54.5\%, the mortality rate associated with Alzheimer's disease was even higher \citep{matias2020death}. The molecular interplay involving overlapping relationships between COVID-19 and other diseases such as carcinomas, neoplasms and Alzheimer's disease is not yet fully understood \citep{al2020impact}.  Given the global spread of COVID-19, it is vital to investigate biological processes and pathways that are common between COVID-19 and other diseases and comorbid conditions. Such advances would ultimately contribute towards better therapeutic outcomes.

Protein-protein interactions (PPIs) are generally modeled using network centrality approaches. Such approaches have been used to understand various disease mechanisms and comorbidities of SARS-CoV-2 in HIV and other immuno-compromised diseases \citep{ambrosioni2021overview}. In the current study, we constructed a disease-gene network which maps (human) genes associated with diseases as well as viral proteins of the Coronaviridae family. 

The aim of the current study is to perform an integrated network analysis between diseases and genes to discover biological processes and pathways shared between COVID-19 and various diseases.

\section{Materials and Methods}
\label{sec:1}

The methodology followed in this study uses network analysis tools for community detection. We constructed a disease-gene network for analysis to identify crucial genes followed by Louvain method based community detection to identify biological process and pathways affected by SARS-CoV-2 infection.

\subsection{Disease-gene Network Construction}
DisGeNET is a comprehensive database comprising of genes and variants associated with human diseases \citep{pinero2021disgenet}. It integrates data from expert-curated repositories, GWAS databases, animal models and the scientific literature. The IntAct Coronavirus data set includes interaction data from the high-throughput multi-level proteomics studies on SARS-CoV-2, SARS and other members of the Coronaviridae family. It contains molecular interactions involving human and viral proteins from the Coronaviridae family, along with a certain proportion of other model organisms.

The disease-gene network upon which the analysis was performed was constructed using IntAct Coronavirus data set that is annotated through DisGeNET \citep{pinero2021disgenet}. The IntAct Coronavirus data set was derived from the IntAct Molecular Interaction database which is a curated database for molecular interaction data \citep{hermjakob2004intact}. For annotation through DisGeNET, only human proteins which interact with members of the Coronaviridae family were taken into account. The disease-gene network was constructed \citep{shannon2003cytoscape} as an undirected biparted graph $G$ ($V_G$, $E_G$), where the 11,301 nodes ($V_G$) represented 828 genes and 10,473 diseases, and the 70,210 edges ($E_G$) represented disease-gene interactions. The network is a biparted graph as all edges connect the set of diseases with the set of human genes, with each edge representing an association from the IntAct Coronavirus data set.

\subsection{Network Analysis}
The degree of a node is defined as the number of edges connected to it. It is one of the measures of network connectivity and network centrality measures used in protein-protein interactions \citep{jeong2001lethality}. Nodes with high degree represent hubs. Figure 1 \label{N_C_D} shows the number of nodes and their respective degrees. Functional gene enrichment analysis was performed to reveal critical pathways that could be targeted for clinical purposes.

\begin{figure}[h!]
\centering
\includegraphics[width=3.0in]{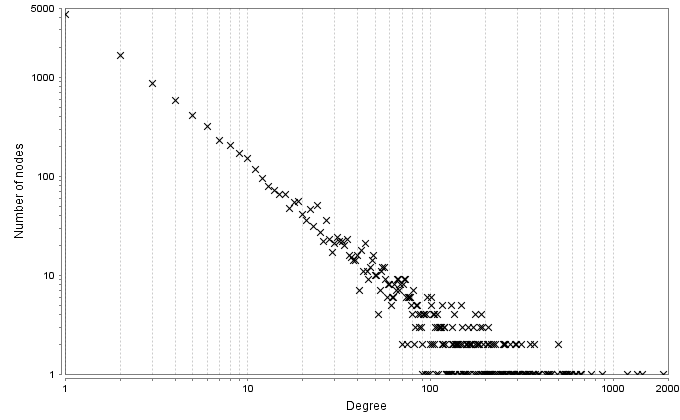}
\caption{Very high degrees for only a fraction of nodes shows that only a smaller set of genes and diseases have high \textit{interactions} with other nodes. For instance, VEGFA is known to be involved in 1899 diseases, while malignant neoplasms are associated with 459 genes. All the genes, taken from IntAct Coronavirus data set \citep{pinero2021disgenet}, are known to be associated with SARS-CoV-2 too.}
\label{N_C_D}
\end{figure}

\subsection{Network Modularity}

Modularity is defined as the fraction of edges that fall within the given groups minus the expected fraction if edges were distributed at random \citep{brandes2008robert}. The modularity for undirected graphs lies in the range of ${\displaystyle (-1/2, 1)}$. A positive modularity indicates that the number of edges within a community exceeds the expected value by random chance. Modularity (Louvain method) of the disease-gene network was 0.28 among all the 22 classes signifying a positive modular structured network (see Figure 2\label{N_M}).

\begin{figure}[h!]
\centering
\includegraphics[width=3.2in]{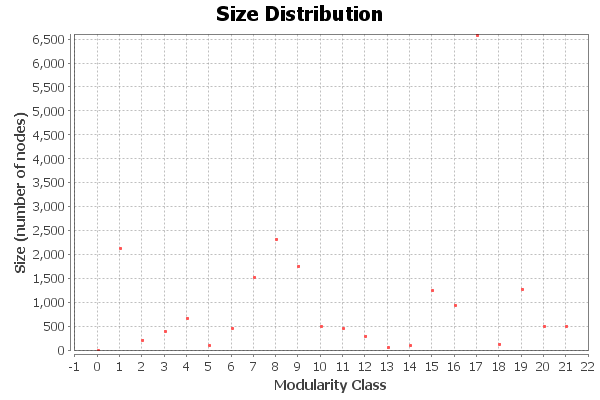}
\caption{The network modularity of the disease-gene network constructed was 0.28 signifying a positive modular structured network. Shown here are sizes (number of nodes) of the 22 modularity classes obtained.}
\label{N_M}
\end{figure}

\subsection{Community Detection}
\vspace{0.1cm}

Community detection algorithms aim to identify the modules and, possibly, their hierarchical organization, based only on the information encoded in the graph topology \citep{fortunato2010community}. The algorithms assign nodes to communities based on the hierarchical organization of the structure of the network \citep{ashourvan2019multi}. Community detection was successfully applied in protein function annotation, identification of disease modules, disease gene prediction and studies on drug-targeted therapies \citep{singhal2020multiscale, gulbahce2008art}. 

Community Detection Application and Service framework (CDAPS) is a framework that integrates identification, annotation, visualization and analysis of large-scale multi-scale network communities \citep{singhal2020multiscale}. Genes underlying the same phenotype tend to interact with each other and form network communities \citep{cowen2017network}. Hierarchical community Decoding Framework (HiDeF) is one of the frameworks that proved its utility in identifying robust structures in omics data \citep{zheng2021hidef}. It is capable of analyzing the multiscale organization of diverse biological systems. Hence, we used HiDeF (through CDAPS) to gain insights into the hierarchical communities present in the disease-gene network.

Community detection was performed using the Louvain algorithm (in HiDeF framework) with a maximum resolution parameter of less than 50 and p-values representing the significance \citep{sham2014statistical}. p-value is defined as the probability of discovering at least \textit{n} genes with the annotation (GO:BP), in a community module with \textit{n} genes.

The community persistence threshold was kept at 5 to remove the unstable clusters. This resulted in a hierarchical network with communities as nodes and their hierarchical relationships as edges.  We identified significant communities to study network connectedness and the number of hierarchical modules of the overall network. For functional enrichment, the communities were based on the significance of overlap with gene sets from curated gene ontology databases. The Louvain method based on network modularity (Q) is defined as  \citep{blondel2008fast}:

\begin{eqnarray}
{\displaystyle Q={\frac {1}{2m}}\sum \limits _{ij}{\bigg [}A_{ij}-{\frac {k_{i}k_{j}}{2m}}{\bigg ]}\delta (c_{i},c_{j})}
\end{eqnarray}

where ${\displaystyle A_{ij}}$ represents the edge weight between nodes ${\displaystyle i}$ and ${\displaystyle j}$;
${\displaystyle k_{i}}$ and ${\displaystyle k_{j}}$ are the sum of the weights of the edges attached to nodes ${\displaystyle i}$ and ${\displaystyle j}$ respectively; ${\displaystyle m}$ is the sum of all of the edge weights in the graph; ${\displaystyle c_{i}}$ and ${\displaystyle c_{j}}$ are the communities of the nodes; and ${\displaystyle \delta }$  is Kronecker delta function $( {\displaystyle \delta (x,y)=1}$ if ${\displaystyle x=y}$, ${\displaystyle 0}$ otherwise).

\begin{algorithm}[h!]
\bigskip
\SetAlgoLined
\SetKwRepeat{REPEAT}{repeat}{until}
\SetKwFor{IF}{if}{then}{endif}
\SetKwFor{FOR}{for}{do}{endfor}
\SetKwFor{WHILE}{while}{do}{endwhile}
G the initial network\\
\REPEAT{}{
{Put each node of G in its own community}\;
\WHILE {some nodes are moved}{
\FOR {all node n of G}{
place n in its neighboring community\\ including its own which maximizes\\ the gain in modularity
}
}
\IF {the new modularity is greater than the initial}{
G = the network between communities of G\;
\Else{Terminate}}
}
\caption{Louvain Algorithm for Community Detection}
\label{L_A}
\end{algorithm}

\subsection{Functional Pathway Enrichment Analysis}
g:Profiler is a widely used tool to detect statistically significant gene ontology terms that highlight biological processes enriched in gene lists with known functional information \citep{raudvere2019g}. Using g:Profiler, genes were mapped to known functional annotations to detect statistically significant enriched pathways for functional enrichment (see Table 2 \label{tbl:biological_process}). This process allows over-representation analysis that highlights statistically significant pathways.

We selected all nodes with minimum Jaccard index value \citep{jaccard1912distribution} for overlap greater than 0.05 and the range of p-value \citep{storey2003statistical} upto a maximum of 0.00001. The information for functional enrichment is fetched from widely known databases such as KEGG database \citep{kanehisa2000kegg}, WikiPathways \citep{pico2008wikipathways}, Reactome pathway database \citep{joshi2005reactome} and other widely known databases for gene ontology.

\begin{figure*}
\centering
\includegraphics[width=6.5in]{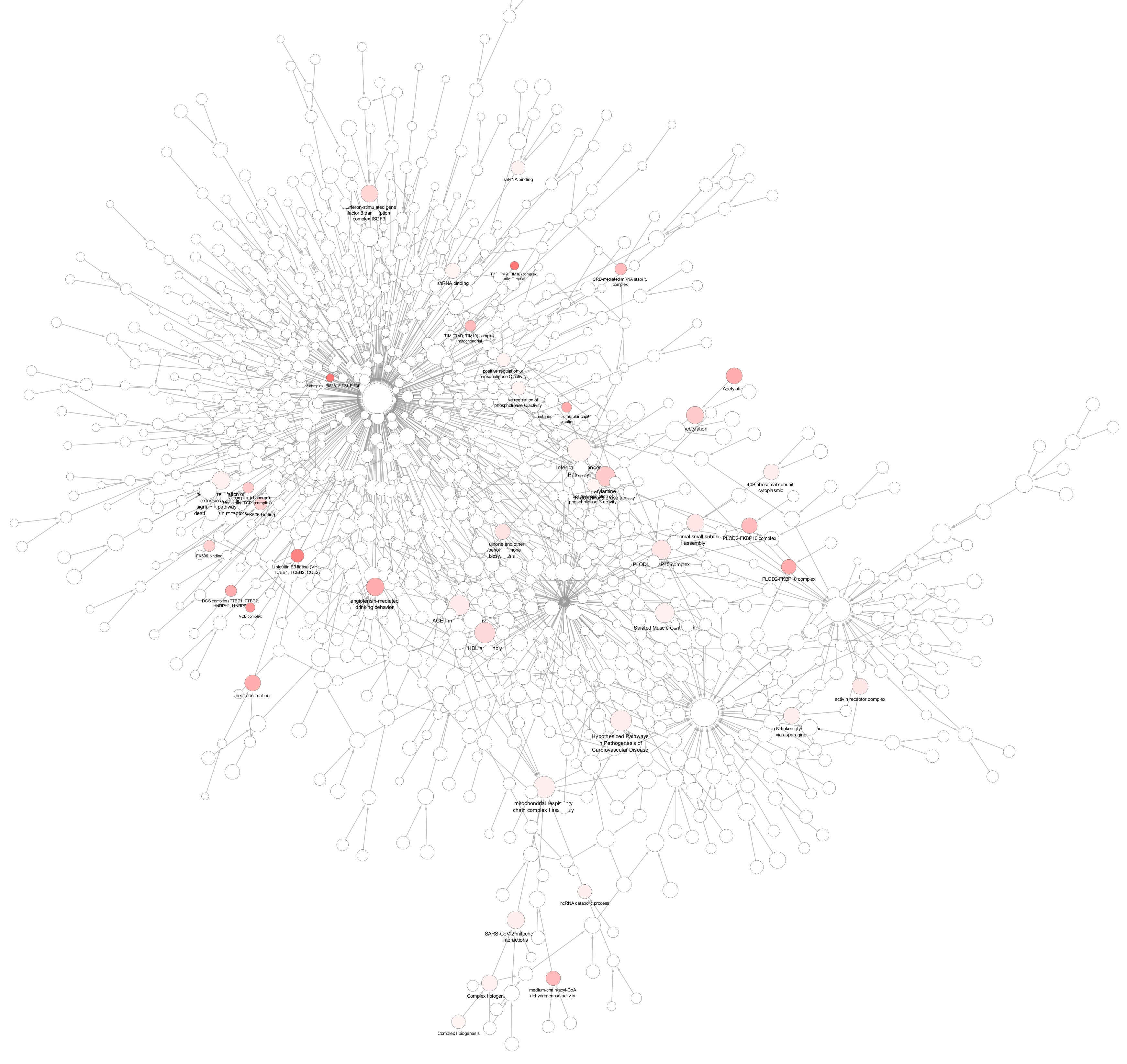}
\caption{Network representation of the most significant biological processes and pathways. Community detection (Louvain method) was executed using Hierarchical community Decoding Framework. The associations of functional enrichment are represented as a hierarchical network for community detection using over-represented analysis wherein the size of a node represents its significance in the Disease-Gene Network.}
\label{N_C_D}
\end{figure*}

\begin{table*}[ht]
	\begin{tabular}{c@{\hspace{2cm}} c}
		(a) High degree genes & (b) High degree diseases \\
		\begin{tabular}{lll}
			\hline \\
			Rank & Gene & Degree \\
			\hline\\
			1 & VEGFA 	& 1899 \\
			2 & BCL2 	& 1456 \\
			3 & CTNNB1 	& 1368 \\
			4 & ALB		& 1198 \\
			5 & COX2 	& 875  \\
			6 & AGT 		& 765  \\
			7 & HLA-A 	& 672  \\
			8 & HMOX1 	& 666  \\
			9 & FGF2 	& 635  \\
			10 & COMT	& 622  \\
			11 & MET 	& 594  \\
			12 & PLG 	& 586  \\
			13 & PCNA 	& 581  \\
			14 & PARP1 	& 565  \\
			15 & AIMP2 	& 555  \\
			16 & FBN1 	& 552  \\
			17 & HSPA4 	& 550  \\
			18 & GNAS 	& 536  \\
			19 & STAT1 	& 531  \\
			20 & AHSA1 	& 526  \\
			21 & TGFBR2	& 502  \\
			22 & HSPA1B	& 502  \\
			23 & DNMT1 	& 496  \\
			24 & HSPA1A 	& 458  \\
			25 & DPP4 	& 451  \\    
			\hline\\
		\end{tabular}
		&
		\begin{tabular}{lll}
			\hline \\
			Rank & Disease Name & Degree \\
			\hline\\
			1 & Neoplasms (unclassified)				& 535 \\
			2 & Malignant Neoplasms 					& 459 \\
			3 & Primary Malignant Neoplasms 			& 444 \\
			4 & Malignant Neoplasm of breast 		& 372 \\
			5 & Tumor cell Invasion 					& 368 \\
			6 & Breast Carcinoma 					& 363 \\
			7 &  Liver Carcinoma 					& 352 \\
			8 &  Carcinogenesis	 					& 347 \\
			9 &  Neoplasm metastasis 				& 344 \\
			10 & Colorectal carcinoma 				& 309 \\
			11 & Malignant neoplasm of prostrate 	& 284 \\
			12 & Prostrate carcinoma 				& 276 \\
			13 & Malignant neoplasm of lung	 		& 263 \\
			14 & Carcinoma of lung 					& 262 \\
			15 & Primary malignant neoplasm of lung  & 257 \\
			16 & Non-small cell Lung carcinoma 		& 241 \\
			17 & Tumor progression 					& 235 \\
			18 & Alzheimer's Disease 				& 226 \\
			19 & Malignant neoplasm of stomach		& 221 \\
			20 & Stomach carcinoma					& 209 \\
			21 & Malignant neoplasm of colon			& 208 \\
			22 & Glioblastoma						& 198 \\
			23 & Glioblastoma Multiforme				& 195 \\
			24 & Glioma								& 193 \\
			25 & Melanoma							& 188 \\    
			\hline\\
			\end{tabular}
	\end{tabular} 
	\caption{List of high-degree nodes in the Disease-Gene network.}
	\label{tbl:d_o_d}
\end{table*}

\begin{table*}[ht]
	\footnotesize
	\begin{tabular*}{\textwidth}{@{\extracolsep{\fill}}llllll}
		\hline
		Rank & Gene Ontology / Biological process & P-value & -log(P-value) & Reference \\
		\hline
		1 & Mitochondrial respiratory chain complex I assembly
		& 8.71E-12 & 11.05998 & GO:0032981\\
		2 & Ubiquitin E3 ligase & 1.63E-10 & 9.787812 & CORUM:622\\
		3 & Activin receptor complex & 2.82E-10 & 9.549751 & GO:0048179\\
		4 & Arylamine N-acetyltransferase activity
		& 1.49E-09 & 8.826814 & GO:0004060\\
		5 & Complex I biogenesis & 1.49E-08 & 7.826814 & REAC:R-HSA-6799198\\
		6 & FK506 binding & 3.23E-08 & 7.490797 & GO:0005528\\
		7 & HDL assembly & 5.21E-08 & 7.283162 & REAC:R-HSA-8963896\\
		8 & Integrated Cancer Pathway & 3.76E-07 & 6.424812 & WP:WP1971\\
		9 & ACE Inhibitor Pathway & 5.50E-07 & 6.258848 & WP:WP554\\
		10 & Hypothesized Pathways in Pathogenesis of CVD & 5.72E-07 & 6.242604 & WP:WP3668\\
		11 & 40S ribosomal subunit, cytoplasmic & 7.42E-07 & 6.129596 & CORUM:305\\
		12 & TIM complex, mitochondrial & 8.17E-07 & 6.087778 & CORUM:623\\
		13 & EIF3 complex & 8.17E-07 & 6.087778 & CORUM:4399\\
		14 & VCB complex & 8.25E-07 & 6.083546 & GO:0030891\\
		15 & Acetylation & 1.09E-06 & 5.962574 & REAC:R-HSA-156582\\
		16 & SARS-CoV-2 mitochondrial interactions & 1.23E-06 & 5.910095 & WP:WP5038\\
		17 & snRNA binding & 1.30E-06 & 5.886057 & GO:0017069\\
		18 & Complex I biogenesis & 1.42E-06 & 5.847712 & REAC:R-HSA-6799198\\
		19 & ribosomal small subunit assembly & 1.50E-06 & 5.823909 & GO:0000028\\
		20 & PLOD2-FKBP10 complex & 1.65E-06 & 5.782516 & CORUM:7000\\
		21 & negative regulation of extrinsic apoptotic signaling pathway & 1.73E-06 & 5.761954 & GO:1902042\\
		22 & CRD-mediated mRNA stability complex & 2.06E-06 & 5.686133 & GO:0070937\\
		23 & TIM complex, mitochondrial & 2.06E-06 & 5.686133 & CORUM:623\\
		24 & ncRNA catabolic process & 2.47E-06 & 5.607303 & GO:0034661\\
		25 & Acetylation & 2.89E-06 & 5.539102 & REAC:R-HSA-156582\\
		26 & angiotensin-mediated drinking behavior & 3.28E-06 & 5.484126 & GO:0003051\\
		27 & positive regulation of phospholipase C activity & 4.44E-06 & 5.352617 & GO:0010863\\
		28 & DCS complex & 4.90E-06 & 5.309804 & CORUM:1288\\
		29 & snRNA binding & 6.58E-06 & 5.181774 & GO:0017069\\
		30 & Ubiquinone and other terpenoid-quinone biosynthesis & 6.71E-06 & 5.173277 & KEGG:00130\\
		31 & metanephric glomerular capillary formation & 7.31E-06 & 5.136083 & GO:0072277\\
		32 & CCT complex & 7.62E-06 & 5.118045 & CORUM:126\\
		33 & heat acclimation & 8.01E-06 & 5.096367 & GO:0010286\\
		34 & Striated Muscle Contraction & 8.70E-06 & 5.060481 & REAC:R-HSA-390522\\
		35 & protein N-linked glycosylation via asparagine & 9.63E-06 & 5.016374 & GO:0018279\\
		36 & medium-chain-acyl-CoA dehydrogenase activity & 9.84E-06 & 5.007005 & GO:0070991\\
		37 & Interferon-stimulated gene factor 3 transcription complex & 9.90E-06 & 5.004365 & CORUM:60\\ 
		\hline
	\end{tabular*}
\caption{List of significantly enriched biological processes and pathways along with their community size. The number of genes involved are less than 7 and the corresponding p-value threshold is less than 9.9E-06.}
	\label{tbl:biological_process}
\end{table*}

\begin{table*}[!htbp]
	\begin{tabular*}{\textwidth}{@{\extracolsep{\fill}}lllll}
		\hline
		Rank & Gene Ontology / Biological process & G$_c$ & FDR \\
		\hline
		1 & Mitochondrial respiratory chain complex I assembly & 7 & 9.60E-14\\
		2 & Ubiquitin E3 ligase & 3 & -\\
		3 & Activin receptor complex & 4 & 5.84E-11\\
		4 & Arylamine N-acetyltransferase activity & 3 & -\\
		5 & Complex I biogenesis & 4 & 1.62E-07\\
		6 & FK506 binding & 3 & 0.0007\\
		7 & HDL assembly & 3 & 0.0211\\
		8 & Integrated Cancer Pathway & 4 & 2.16E-08\\
		9 & ACE Inhibitor Pathway & 3 & 6.19E-07\\
		10 & Hypothesized Pathways in Pathogenesis of CVD & 3 & 1.78E-06\\
		11 & 40S ribosomal subunit, cytoplasmic & 3 & 1.56E-05\\
		12 & TIM complex, mitochondrial & 2 & 0.00034\\
		13 & EIF3 complex & 2 & 0.0012\\
		14 & VCB complex & 2 & 0.00019\\
		15 & Acetylation & 2 & 6.81E-05\\
		16 & SARS-CoV-2 mitochondrial interactions & 3 & 3.55E-06\\
		17 & snRNA binding & 3 & 4.20E-05\\
		18 & Complex I biogenesis & 3 & 5.37E-05\\
		19 & ribosomal small subunit assembly & 3 & 1.37E-05\\
		20 & PLOD2-FKBP10 complex & 2 & -\\
		21 & Negative regulation of extrinsic apoptotic signaling & 3 & 0.00012\\
		22 & CRD-mediated mRNA stability complex & 2 & 0.00025\\
		23 & TIM complex, mitochondrial & 2 & 0.00034\\
		24 & ncRNA catabolic process & 3 & 6.73E-05\\
		25 & Acetylation & 2 & 6.81E-05\\
		26 & angiotensin-mediated drinking behavior & 2 & 0.00040\\
		27 & positive regulation of phospholipase C activity & 3  & 0.00019\\
		28 & DCS complex & 2 & -\\
		29 & snRNA binding & 3 & 4.20E-05\\
		30 & Ubiquinone and other terpenoid-quinone biosynthesis & 2 & 0.00014\\
		31 & metanephric glomerular capillary formation & 2 & 0.00067\\
		32 & CCT complex & 2 & 0.00059\\
		33 & heat acclimation & 2 & -\\
		34 & Striated Muscle Contraction & 3 & 1.47E-05\\
		35 & protein N-linked glycosylation via asparagine & 3 & 5.61E-05\\
		36 & medium-chain-acyl-CoA dehydrogenase activity & 2 & 0.00017\\
		37 & Interferon-stimulated gene factor 3 complex & 2 & -\\ 
		\hline
	\end{tabular*}
	\caption{Functionally enriched biological processes validated using STRING along with the number of genes involved (G$_c$) in the community and the False Discovery Rates (FDR) of each GO:BP node.}
	\label{tbl:fdr}
\end{table*}

\begin{figure*}[h!]
	\centering
	\includegraphics[width=5.0in]{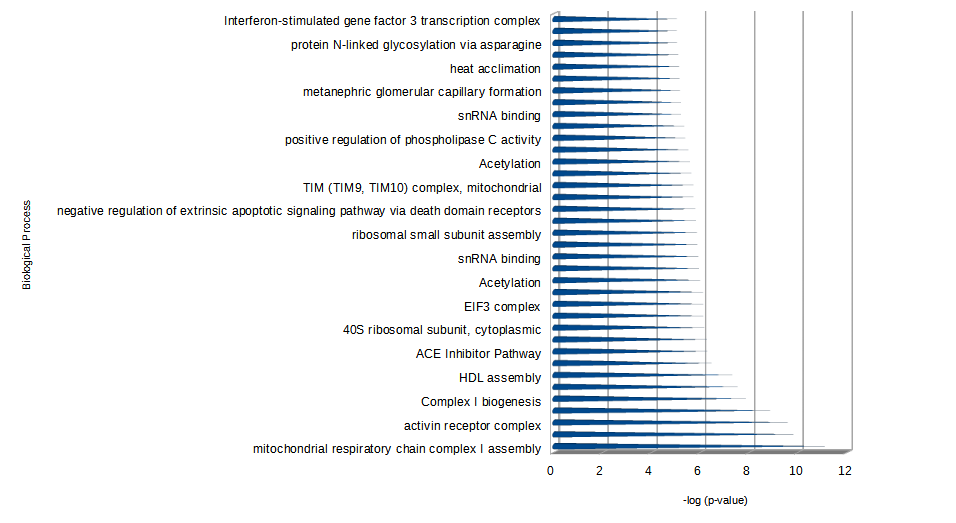}
	\caption{List of significantly enriched biological pathways based on -log(p-value).}
	\label{N_C_D}
\end{figure*}
 
\section{Results and Discussion}

Treating patients with life threatening diseases during COVID-19 pandemic has been quite challenging. There is an urgent need for  treatment strategies that utilize therapies involving immunosuppression. Therapeutic clinical applications used in various immunotherapies have led to the worsening of COVID-19 outcomes among patients suffering from cancer \citep{derosa2020immuno}. Acute respiratory distress syndrome (ARDS), pulmonary embolism, acute myocardial infarction and septic shock were the leading causes of death among cancer patients \citep{addeo2020cancer}. In developing adverse COVID-19 outcomes, cancer patients are at higher risk \citep{turnquist2020cytokine}. The clinical manifestations of neurological disorders, carcinomas and COVID-19 infected patients are to a certain extent interrelated with the modulation of immune system involving cytokine storm and immunosuppression \citep{bakouny2020covid}. It has been hypothesized that patients suffering from the above stated conditions being more vulnerable to viral infections due to compromised immune responses \citep{ferini2021covid}. Therefore, it becomes critical to investigate as to why patients suffering from immunocompromised disorders are at higher risk, and to determine how SARS-CoV-2 infection affects the overall functionality of patients with comorbidities. We present the insights obtained from network analysis and functional enrichment studies on the disease-gene network. 

\subsection{Analysis of Disease-Gene Network}

The disease-gene network was constructed as described in Section 2.1. This network is a biparted graph with edges connecting diseases to functionally connected (human) genes affected by COVID-19. The network comprised of 11,301 nodes (828 human genes and 10,473 diseases) and 70,210 edges. 

As shown in Tables 1 (a-b) \label{tbl:d_o_d}, the top 25 diseases ranked by degree were either neoplasms or carcinomas. The lone exception to this list was Alzheimer's disease, a neurodegenerative disease that affects a large population. 

The top 10 genes that were involved with most diseases were identified as VEGFA, BCL2, CTNNB1, ALB, COX2, AGT, HLA-A, HMOX1, FGT2 and COMT. The degrees were in the range 622 (COMT) to 1,899 (VEGFA). These results reflect the significance of hubs in the overall disease-gene network. 

Escaping immune system mechanism is a prerequisite for tumor development in carcinomas. Angiogenic entities such as VEGFA play a significant role \citep{schreiber2011cancer} in the development of immunosuppressive micro environment \citep{voron2015vegf}. Consequently, strategies involving treatment with anti-VEGF drugs such as bevacizumab besides VEGFR-mediated signaling therapies may deliver favorable outcomes. Studies have already shown that such treatments improve anti-inflammatory responses and oxygen perfusion, culminating in the alleviation of clinical symptoms in severe COVID-19 patients \citep{pang2021efficacy}.

\subsection{Network Topology Community Detection}

We performed community detection on the disease-gene network using Louvain algorithm \citep{blondel2008fast} to obtain 1,052 significant communities. The p-value was kept as 0.00001. 

\subsection{Functional Enrichment Analysis}

Our main focus was to highlight the critical processes affected by COVID-19 disease. Figure 3 shows the network representation of top-ranked biological processes and pathways obtained through the analysis. To identify significantly enriched biological processes and pathways, the corresponding p-value threshold was kept less than 9.9E-06. As listed in Table 2, we identified 37 such key biological processes and critical pathways affected by Coronaviridae family in the overall disease gene network with p-values ranging between 8.71E-12 to 9.90E-6. 

As shown in Table 3 \label{tbl:fdr}, the maximum number of genes involved in all the identified communities was less than 7.
The top 10 biological processes/critical pathways were associated with the mitochondrial respiratory chain I complex assembly, Ubiquitin E3 ligase, Activin receptor complex, Arylamine N-acetyltransferase activity, Complex I biogenesis, FK506 binding, HDL assembly, Integrated Cancer Pathway and ACE Inhibitor Pathway and Hypothesized Pathways in Pathogenesis of Cardiovascular Diseases (as shown in Figure. 4) \label{tbl:biological_process}.

\subsection{Validation}

For validation, we correlated our findings with STRING database which is a comprehensively covers human proteins in the form of manually curated function protein interaction network \citep{szklarczyk2021string}. As shown in Table 3 \label{tbl:fdr}, 31 out of all the 37 significant communities (84\%) detected were validated (supplementary table attached) with least False Discovery Rate (FDR), demonstrating that the majority of highlighted GO:BP terms are clinically significant. 

\section{Conclusion}

In the current study, we analyzed the disease-gene network linked to SARS-CoV-2 and other members of the Coronaviridae family. Our study strongly suggests that the patients with comorbid conditions, especially various carcinomas, neoplasms and Alzheimer's disease, are the most vulnerable among all the disease classes infected by SARS-CoV-2.

Using network topology and community detection algorithm followed by functional enrichment, the top 10 potential genes that were involved with multiple diseases involving various carcinomas and neoplasms were identified as VEGFA, BCL2, CTNNB1, ALB, COX2, AGT, HLA-A, HMOX1, FGT2 and COMT. We report 37 key biological processes and disease pathways affected by Coronavirus infection in the disease gene network. As an outcome of our study, we make case for clinical investigations towards anti-VEGF therapies in patients suffering from various carcinomas and neoplasms infected with COVID-19. 

%

\bibliographystyle{unsrt}
\bibliography{compbio}   

%
%


\end{document}